\documentclass[11pt]{article}

\newcommand{\uvec}{{\bf u}}

\newcommand{\vvec}{{\bf v}}

\newcommand{\bR}{{\bf R}}

\newtheorem{theorem}{Theorem}[section]
\newtheorem{prop}[theorem]{Proposition}
\newtheorem{lemma}[theorem]{Lemma}
\newtheorem{corollary}[theorem]{Corollary}

\newcommand{\nn}{\nonumber}

\def\bd{\begin{displaymath}}
\def\ed{\end{displaymath}}

\def\eqref#1{(\ref{#1})} 
\def\qed{\hbox{\hskip 6pt\vrule width6pt height7pt depth1pt
    \hskip1pt}\bigskip}

\def\runinend{\enspace}
\def\ackname{Acknowledgement\runinend}%
\def\acknowledgements{\par\addvspace{17pt}\rmfamily
\def\ackname{Acknowledgements\runinend}%
\trivlist\if!\ackname!\item[]\else
\item[\hskip\labelsep
{\bf\ackname}]\fi}%

\begin{document} 
\bibliographystyle{plain} 

\hfill{} 
 
\thispagestyle{empty}

\begin{center}{\bf \Large Low regularity solutions to a gently 
stochastic nonlinear wave equation in nonequilibrium statistical 
mechanics } 
\vspace{5mm}

\large{ \bf Luc  Rey-Bellet}

{\small\it Department of Mathematics and Statistics, 
University of Massachusetts, \\ 
Amherst, MA 01003, Email: lr7q@math.umass.edu \\ }
\vspace{5mm}

\large{\bf Lawrence E. Thomas}  \\    

{\small\it Department of Mathematics, University of Virginia, \\
Charlottesville, VA 22903, Email: let@virginia.edu \\ }

\end{center}

\setcounter{page}{1} 
\begin{abstract} We consider a system of stochastic partial
differential equations modeling heat conduction in a non-linear
medium. We show global existence of solutions for the system in
Sobolev spaces of low regularity, including spaces with norm beneath
the energy norm. For the special case of thermal equilibrium, we also
show the existence of an invariant measure (Gibbs state).
\end{abstract}

\section{Introduction} In this article we consider the following system 
of partial differential equations 
\begin{eqnarray}
\partial_t\phi(x,t)&=& \pi(x,t)\nonumber\\  
\partial_t\pi(x,t)&=& (\partial_x^2 - 1)\phi(x,t)- \mu \phi^3(x,t)-r(t)  \alpha(x)
\nonumber\\
dr(t)&=&  \left(r (t)-\langle\alpha,\pi(t)\rangle\right)dt + 
\sqrt{2T}d\omega(t)\,. \label{theeq}
\end{eqnarray}
In Eqs.\eqref{theeq} $(\phi,\pi)$ is a pair of scalar fields satisfying 
periodic boundary conditions with $x \in [0,2\pi]$. The vector-valued 
functions $\alpha = (\alpha_{1}, \cdots, \alpha_{K})$ are given with 
$\alpha_j \in H^{\gamma}$, for some $\gamma >0$, for all $j$. 
The vector $r=(r_{1}, \cdots, r_{K})$ is in $ \bR^{K}$, and 
$\omega = (\omega_{1}, \cdots, \omega_{K})$ 
is a $K$-dimensional Brownian motion. 
The matrix $T$ is given by  $T= {\rm diag}(T_{1}, \cdots, T_{K})$, 
and $T_{j}$ will be interpreted as a temperature. The parameter $\mu$ is a 
coupling constant; we will be primarily interested in the cases 
$\mu=0$ (linear Klein-Gordon equation) and $\mu>0$ (nonlinear 
defocusing linear wave equation). 

The system of equations \eqref{theeq} arises from a model for heat
conduction in a nonlinear medium. It can be derived from first
principles from a Hamiltonian system which consists of $K$ linear wave
equations in $\bR$ coupled to a nonlinear wave equation in
$[0,2\pi]$. The total Hamiltonian is given by
\begin{eqnarray} 
H\,&=&\,\sum_{j=1}^{K}  \int_{\bR} \frac{1}{2} 
\left(|\partial_{x} u_{j}(x) |^{2} + |v_{j}(x)|^{2}\right) \, dx  
\nn\\
&& \,\,\,+ \int_{[0,2\pi]} \frac{1}{2}  \left(
|\partial_{x} \phi(x) |^{2} + |\phi(x) |^{2} + |\pi(x)|^{2} 
\right) + \frac{\mu}{4} |\phi(x) |^{4}  \, dx
\nn \\ 
&& \,\,\,+ \sum_{j=1}^{K} \left(\int_{\bR} \partial_{x} u_{j}(x) \rho_{j}(x) \, dx 
\right) 
\left( \int_{[0,2\pi]} \partial_{x} \phi(x) \alpha_j(x) \, dx \right) 
\label{totham}
\end{eqnarray}
One assumes further that the initial conditions of the 
$(u_{j},v_{j})$, $j=1, \cdots, K$  (``the reservoirs'') are 
distributed according to Gibbs measures at temperatures $T_{j}$. 
These measures are (formally) expressed as 
\begin{equation}\label{initial}
Z^{-1} \exp \left( -\frac{1}{2T_{i}} \int_{\bR} 
\left(|\partial_{x} u_{j}(x) |^{2} + |v_{j}(x)|^{2}\right) \, dx 
\right)  \prod_{x \in \bR} du_j{(x) } dv_{j}(x)  \label{lg} \,,
\end{equation}
and they are simply the product of a Wiener measure (for the  fields $u_j$) 
with a white noise measure (for the momenta $v_j$). 

We refer to \cite{EPR1} or \cite{RT2,RB} for details on the derivation
of equations \eqref{theeq} from the Hamiltonian system \eqref{totham}
with initial conditions \eqref{initial}, at least in the case where the
nonlinear wave equation is replaced by a chain of nonlinear
oscillators (formally a discrete wave equation). In that case one
obtains a set of stochastic ordinary differential equations. The
derivation is essentially the same as for the model considered here
and will not be repeated. We simply remark that the derivation of Markovian 
equations is possible due a particular choice of the coupling functions 
$\rho_j$. 

In a series of papers \cite{EPR1,EPR2,RT1,RT2,RT3,EH1,EH2,RB} about
the chain of nonlinear oscillators, the existence, uniqueness, and
strong ergodic properties of invariant measures have been
established. Moreover, a number of properties of these invariant
measures have been elucidated, such as existence of heat flow,
positivity of entropy production, and symmetry properties of
entropy production fluctuations. These invariant measures represent
stationary states which generalize Gibbs distributions to
non-equilibrium situations where there is heat flow. Ultimately our goal 
is to establish similar properties for the systems of equations \eqref{theeq}. 
But we study here the more immediate problems of
existence of global solutions-- a prerequisite for studying the
existence of stationary states-- and existence and invariance 
of an equilibrium (Gibbs) measure.

In the case of equilibrium, that is, when all temperatures are equal,
$T_{j}=T$ for all $j=1, \cdots, K$, we will prove below that 
there is an invariant state given formally by the (non-Gaussian) Gibbs 
measure 
\begin{eqnarray}\label{Gibbsstate}
&&d\nu= 
Z^{-1} \exp \left(-\frac{1}{2T} \int_{[0,2\pi]}\!\!\!\!\!\!\!\! (|\partial_{x} 
\phi(x) |^{2} + |\phi(x) |^{2} + \frac{\mu}{2} |\phi(x) |^{4}  + 
|\pi(x)|^{2} ) \, dx  \right) \nn \\
&& \quad \quad \quad \times \exp\left ( -\frac{1}{2T} r^{2} \right)  
\,dr \,\prod_{x \in [0,2\pi]}  d\phi(x)  d\pi(x) \,. \label{nlg} 
\end{eqnarray}
To make sense of this measure, one considers first the Gaussian
measure $\nu^{0}$ for the case $\mu=0$. Its support is contained in
$H^{s}\times H^{s-1}\times \bR^{K}$ for any $s < \frac{1}{2}$ and,
with probability $1$, $\phi$ is also a continuous function. Hence we
can think of the measure $\nu$ as the measure which is absolutely
continuous with respect to $\nu^{0}$ with a Radon-Nikodym derivative
proportional to $\exp(-\mu\int|\phi(x)|^{4}dx/4T)$.  We {\em expect}, but
have by no means proved, that the invariant measure for different
temperatures, if one exists, has similar support properties.  But with
this intuition, it is appropriate to seek solutions of \eqref{theeq}
in spaces of rough data $H^{s} \times H^{s-1} \times \bR^{K}$ with $s
< \frac{1}{2}$.  Indeed we show the global existence of strong
solutions, for $1/3\leq s< 1$ (see Corollary  \ref{corollary} and the
remark following it).  We believe that these spaces, with at least
$1/3\leq s <1/2$, are natural to the invariant measure problem.

Clearly, in these spaces no energy conservation (or bounds on the
energy growth/dissipation) is available. In recent years, however,
Bourgain \cite{Bo1}, Keel and Tao \cite{KeTa} and many others have
developed techniques to show global existence for wave equations and
other Hamiltonian PDE's in Sobolev spaces below the energy norm.  A
review of recent results with an extensive bibliography can be found
in \cite{CKSTT}.  Here, we use and extend these methods to
establish global existence of solutions for wave equations coupled to
heat reservoirs, i.e. with noise and dissipation.

In the last section, we show that solutions to an ultra-violet cut-off
version of our system of equations, Eq.(\ref{theeq}), converge as the
cut-off is removed.  This result is then applied to show that the
equilibrium Gibbs state $\nu$ described above, Eq.(\ref{Gibbsstate}),
is indeed an invariant measure in the case of equilibrium.  Note that
Gibbs measures for nonlinear wave equations (and nonlinear
Schr\"{o}dinger equations) have been constructed and studied by
several authors, (Lebowitz, Rose and Speer \cite{L}, Zhidkov
\cite{Zh}, McKean and Vaninsky \cite{Mc}, Bourgain \cite{Bo2,B},
Brydges and Slade \cite{BS}) but for isolated systems only. Note that
in these works Gibbs measures for any temperature are invariant while
in our case the temperature is selected by the coupling to the
reservoir. Our work is also related in spirit to various recent works
on the ergodic properties of randomly forced dissipative equations,
see e.g. \cite{BKL,EMS,EH3,KS,MY} and others. The main and very
important differences are that our equation is Hamiltonian rather
than parabolic so that there is no intrinsic smoothing in the
equations, and that the dissipation is very weak.

\subsection{Notation}

It is convenient to write our system as Bourgain does \cite{Bo1}. Set
\begin{equation}
     u= \phi+ \frac{i}{B}\pi
\end{equation}
where $B$ is the operator defined $B=\sqrt{-\partial_x^2+1}$.  Note that
$\phi= {\Re u}$ and $\frac{1}{B}\pi= {\Im u}$ are respectively the
real and imaginary parts of $u$. Thus our differential equations can
be written,
\begin{eqnarray}\label{DIFFEQ}
   i\partial_tu&=& Bu +\frac{1}{B} (\mu\phi^3+r \alpha)\nonumber\\
 dr (t)&=& - (r(t)-\langle\alpha,\pi(t)\rangle)\,dt +\sqrt{2T}\,d\omega(t)
\,.
\end{eqnarray}
Let
 \begin{equation}
 \uvec(\omega,t) = (u,r)=
(\phi+\frac{i}{B}\pi, r),
\end{equation}
and let $\uvec_o(\omega,t)= (u_o,r_o)(\omega,t)$ be the
corresponding solution to the differential equations but with the
non-linearity turned off, $\mu = 0$.

For a vector quantity $\uvec= (u,r)$,
we introduce the norms
\begin{equation}
    \|\uvec\|_{H^s}= (r^2+\|u\|_{H^s}^2)^{1/2}.
    \end{equation}
The energy of a vector $\uvec$ is defined by
\begin{equation}\label{energydef}
   {\cal E}(\uvec)= \frac{1}{2}\|u\|_{H^1}^2+\frac{1}{2}r^{2} 
+\frac{\mu}{4}\int(\Re u)^{4}dx.
\end{equation}

\section{Estimates for the Linear Wave Equation}

In this section we collect basic estimates for the linear system,
$\mu=0$.  These estimates actually establish global existence for
this system.

The first step is to consider the linear (dissipative) system without
the random driving terms,
\begin{eqnarray}\label{ls}
     \frac{du_o}{dt} &=&-iBu_o -i\frac{1}{B}\alpha  r_o\nonumber\\
     \frac{dr_o}{dt} &=& \langle B\alpha, \Im u_o\rangle -r_o
     \end{eqnarray}
where $\Im u_o$ is the imaginary part of $u_o$. Set $L_o=
B_o+P$ with
\begin{eqnarray}\label{ls2}
B_o&=& \left(\begin{array}{cc}
        -iB& 0\\
        0& -1
        \end{array}\right)\nonumber\\
P&=& \left(\begin{array}{cc}
         0&-i\frac{1}{B}|\alpha\rangle\\
         \langle B\alpha|\Im &0
         \end{array} \right)
\end{eqnarray}
Symbolically, the solution of this system Eqs.(\ref{ls}) is given by
$e^{tL_o}\uvec_o(0)$ with $L_o= B_o+P$. The system should be regarded
as linear in a function space of complex functions over the reals (so
that $\Im$ is linear).

\begin{lemma}\label{BSlemma}
Assume that $\alpha\in H^{\gamma}$ for some $\gamma>0$ and
$0\leq s<1$. For $\lambda_0$ sufficiently large depending on the
$\alpha$'s only, $(B_o+\lambda_0)(L_o+\lambda_0)^{s-1}
(B_o+\lambda_0)^{-s}$ acting in $L^2\oplus {\bf R}$ is defined as a bounded invertible operator. 
\end{lemma}
  
\noindent {\em Proof:}  We have the following operator estimates
(the operators acting in $L^2\oplus {\bf R}$):
  \begin{eqnarray}
    \left \|P\frac{1}{B_o+\lambda_0+\lambda}\right\|&\leq& \frac{c(\lambda_0)}{(1+\lambda)^{\gamma}}\nonumber\\
     \left\|P\frac{1}{(B_o+\lambda_{0}+\lambda) (B_o+\lambda_0)^s}\right\|&\leq&
     \frac{c(\lambda_0)}{(1+\lambda)^{\gamma'}}
     \end{eqnarray}
with $\gamma'= \min\{\gamma+s,1\}$ and with $c(\lambda_0)\rightarrow0$
for $\lambda_0\rightarrow\infty$.  By expanding the resolvent for
$L_{o}$ in a geometric series, convergent for $c (\lambda_0)<1$, one
finds from these estimates that
   \begin{eqnarray}\label{auxB1}
  \lefteqn{(B_o+\lambda_0) (L_{o}+\lambda_0+\lambda)^{-1} (B_o+\lambda_0)^{-s}}\nonumber\\
&=& (B_o+\lambda_0)^{1-s}(B_{o}+\lambda_0+\lambda)^{-1}+{\cal O} \left(\frac{c (\lambda_0)}{(1+\lambda)^{\gamma'}}\right) .
\end{eqnarray}  
Using 
\begin{equation}
    (L_o+\lambda_0)^{s-1}= c_{s}\int_0^{\infty}\frac{d\lambda}{(L_o+\lambda_0+\lambda)\lambda^{1-s}}
\end{equation}
with $c_{s}$ a suitable normalizing constant, and integrating the
previous equation, we obtain
\begin{equation}
(B_o+\lambda_0) (L_{o}+\lambda_0)^{s-1} (B_o+\lambda_0)^{-s}= 1 +{\cal
  O} (c (\lambda_0)),
\end{equation}
which clearly is bounded.  
By choosing $\lambda_0$ large so that $c (\lambda_0)$ is sufficiently
small, we see that $(B_o+\lambda_0) (L_{o}+\lambda_0)^{s-1}
(B_o+\lambda_0)^{-s}$ is invertible. \hfill \qed

\begin{lemma}\label{lemmaBS2} Assume $\alpha \in H^{\gamma}$, with
  $\gamma>0$, $0<s\leq 1$.  
There is a constant $c_3$ depending only on $s$ and the $\alpha$'s,
 such that 
 \begin{equation}\label{aux6c}
\|e^{tL_o}\uvec(0)\|_{H^s}\leq c_3 \|\uvec(0)\|_{H^s}
\end{equation}
for all time $t$.
\end{lemma}

\noindent{\em Proof:} We have that
\begin{equation}
  {\cal{E}}_o(\uvec)\equiv \frac{1}{2}\left(\|u\|_{H^1}^2+
  r^2\right)
  \end{equation}
  is a (degenerate) Liapunov function for the linear system
  Eq.(\ref{ls}). The lemma follows if we can show that for a suitably
  large constant $\lambda_0$, ${\cal{E}}_o((
  L_o+\lambda_0)^{s-1}\uvec)$ is equivalent to $\|(
  B_o+\lambda_0)^s\uvec\|_2^2$, which is in turn equivalent to
  $\|\uvec\|_{H^s}^2$. This is certainly the case if $s=1$.  For
  $s<1$, this amounts to showing that
  $(B_o+\lambda_0)(L_o+\lambda_0)^{s-1} (B_o+\lambda_0)^{-s}$ is a
  bounded invertible operator, which is the content of the previous
  lemma.\hfill \qed

We now provide an estimate for the linear stochastic evolution $\uvec_o$. 

\begin{lemma}\label{lemmasigma0} Assume that $\gamma>0$, $0<s\leq 1$
  and set $\|\uvec_o(0)\|_{H^s}=\beta$.  
There exist constants $c$ and $C$, such that for
$\lambda\geq c\beta$,
\begin{equation}\label{sigma0bound}
    P\left\{\sup_{t'<t}\|\uvec_o(t)\|_{H^s}\geq \lambda \right\}\leq
    C\exp{\left(-\frac{(\lambda-c\beta)^2}{c^2t(1+t)^2}\right)}.
    \end{equation}
    \end{lemma}

\noindent {\em Remarks}: The estimate is certainly not optimal.  It
does not account for the rapid dissipation of energy for small $k$ modes of
$\uvec_o$. The lemma provides a {\em global bound} on the
linear evolution, showing that it doesn't blow up, almost surely.\\

\noindent {\em Proof}:  The solution $\uvec_o(t)$ can be written
in integral form,
\begin{eqnarray}\label{lineq1}
   \uvec_o (t) &=& \int_0^t e^{(t-t')L_{o}} \vvec_o\,d\omega(t') +
   e^{tL_{o}}\uvec_{o}(0)\nonumber\\
&=&  \vvec_o\omega(t)+\int_0^t e^{(t-t')L_{o}}L_{o} \vvec_o\omega(t') dt'+ e^{tL_{o}}\uvec_{o}(0)
   \end{eqnarray}
by integration by parts, with $L_{o}$ defined as in the deterministic case, Eq.(\ref{ls2}), and 
\begin{equation}
\vvec_o= \left(\begin{array}{c}
         0\\ \sqrt{2T}
      \end{array}\right).
\end{equation}
By Lemma \ref{lemmaBS2}, there is a constant $c_{3}$ such that 
$\|e^{tL_o}\uvec_o (0)\|_{H^{s}}   \leq c_{3}\beta$, and
$\|e^{(t-t')L_o}\vvec_o\|_{H^{s}}\leq c_{3}\|\vvec_o\|_{H^{s}}$ (which
is finite) so that from the integral equation Eq.(\ref{lineq1})
above, we obtain the estimate
\begin{eqnarray}
\lefteqn{\|\uvec_o(t)\|_{H^{s}}}\nonumber \\ 
&\leq&\| \vvec_o\|_{H^s}|\omega(t)|+c_{3}t\| L_{o} \vvec_o\|_{H^s}\sup_{t'\leq t}|\omega (t')| +c_{3}\|\uvec_{o}(0)\|_{H^s}. 
\end{eqnarray}
Thus we can write for a suitable constant $c$ that
\begin{equation}
\| \uvec_o(t)\|_{H^s}\leq c( 1+ t) \sup_{t'\leq t}|\omega(t')| +
c\beta,
\end{equation}
with $\beta$ the $H^s$ norm of the initial data $\uvec_o(0)$.

Now if at some time $t'$, with $t'\leq t$, we have that
$\|\uvec_o (t')\|_{H^s}> \lambda$, then evidently
$\frac{\lambda-c\beta}{c(1+t)}\leq \sup_{t'\leq t}|\omega(t')|$,
and so, for $\lambda>c\beta$,
\begin{eqnarray}
P\left\{\sup_{t'\leq t}\|\uvec_o(t')\|_{H^s}> \lambda\right\}&\leq&
P\left\{\sup_{t'\leq t}|\omega(t')|>
\frac{\lambda-c\beta}{c(1+t)}\right\}\nonumber\\
&\leq & 2P\left\{|\omega(t)|> \frac{\lambda-c\beta}{c(1+t)}\right\}\nonumber\\
&\leq &
C\exp{\left(-\frac{(\lambda-c\beta)^2}{c^2t(1+t)^2}\right)}
\end{eqnarray}
for yet another suitable constant $C$ depending on the dimension
of $r$. This concludes the proof of the lemma.\hfill \qed

For later use, we also note here some simple Sobolev inequalities, all in
one-dimension only.

\begin{lemma}\label{Sob}
For $s> (1/2-1/p)$ and $p\geq 2$ there is a constant $c= c (s,p)$ such that
\begin{equation}\label{Sobolev1}
   \|\phi\|_p\leq c\|\phi\|_{H^s}.
   \end{equation}
Also, for $0\leq \theta \leq (1-1/p)$, there is a constant $c$
such that
  \begin{equation}\label{Sobolev2}
  \|\phi\|_{\infty}\leq c\|\phi\|_{2(p-1)}^{\theta}
  \|\phi\|_{H^1}^{1-\theta}.
  \end{equation}
Finally, for $s>1/6$ and $s'\leq \min(0,3s-1)$, or $s'=0$ and $s>1/3$,
there is a constant $c= c(s,s')$
such that for $\phi_1,\,\phi_2,\,\phi_3\in\,\,H^s$,
   \begin{equation}\label{Sobolev3}
  \|\phi_1\phi_2\phi_3\|_{H^{s'}}\leq c \|\phi_1\|_{H^s}\|\phi_2\|_{H^s}\|\phi_3\|_{H^s}.
\end{equation}
 \end{lemma}

\noindent {\it Remark}: The first inequality of the lemma actually
holds with $s= 1/3$ and $p= 6$, as can be proved using the
Hardy-Littlewood-Sobolev inequality.  For convenience we will use this
inequality as well, although it isn't essential for our purposes. But
as a consequence of this remark, the last inequality
Ineq.(\ref{Sobolev3}) holds for $s'=0,\,\,s=1/3$.\\

\noindent{\em Proof:} The first inequality of the lemma is proved
by estimating
\begin{eqnarray}\label{extra}
   \|\hat{\phi}\|_{\ell^{p'}}^{p'}&=& \sum
   _n\frac{1}{(1+n^2)^{sp'/2}}(1+n^2)^{sp'/2}|\hat{\phi}|^{p'}(n)\nonumber\\
&\leq & \|(1+n^2)^{-sp'/2}\|_{\ell^r}\|\phi\|_{H^s}^{p'}
\end{eqnarray}
with $p'$ conjugate to $p$ and $r= 2/(2-p')$, all of which is bounded provided that
$sp'r>1$, i.e., $s> \frac{1}{2}-\frac{1}{p}$.  One then applies
Hausdorff-Young to obtain the first assertion of the lemma.

The second inequality  of the lemma is shown by first noting that
\begin{equation}
      \phi(x)= \frac{1}{2\pi}\sum_n
      \frac{e^{inx}}{(1+n^2)^{1/2}}(1+n^2)^{1/2}\hat{\phi}(n)
\end{equation}
which by the Schwarz inequality gives the special case
($\theta=0$)
\begin{equation}\label{aux1}
\|\phi\|_{\infty}\leq c \|\phi\|_{H^1}.
\end{equation}
  Also, we have that
\begin{equation}\label{aux3}
   |\phi|^p(x)\leq p\int_y^x|\phi|^{p-1}|\phi'(t)|dt+ |\phi|^p(y).
   \end{equation}
Estimating the integral by
$\|\phi\|_{2(p-1)}^{p-1}\|\phi\|_{H^1}$ and then integrating
this inequality (\ref{aux3}) with respect to $y$ over $[0,2\pi]$,
we get
\begin{equation}
2\pi|\phi|^p(x)\leq 2\pi p \|\phi\|_{2(p-1)}^{p-1}\|\phi\|_{H^1}+
\|\phi\|_{2(p-1)}^{p-1}\|\phi\|_2,
\end{equation}
so that
\begin{equation}\label{aux4}
  \|\phi\|_{\infty}\leq c
  \|\phi\|_{2(p-1)}^{1-1/p}\|\phi\|_{H^1}^{1/p}.
\end{equation}
The second inequality of the lemma is then obtained by
interpolation between Ineqs.(\ref{aux1},\ref{aux4}).

To prove the last inequality of the lemma, Ineq.(\ref{Sobolev3}), we
suppose each of the $\phi_i$'s is in $ H^s$ with $s>1/6$.  Pick $p'$
with $s> \frac{1}{p'}-\frac{1}{2}$ and, for later purposes,
$\frac{6}{5}< p'\leq \frac{3}{2}$. By Ineq.(\ref{extra}) above, each
$\hat{\phi}_i$ is in $\ell^{p'}$, and the double convolution
$\hat{\phi}_1*\hat{\phi}_2*\hat{\phi}_3$ is in $\ell^r$ for
$\frac{1}{r}= \frac{3}{p'}-2$ by Young's inequality. Note that $2<
r\leq \infty$.  It is then easy to check that
$(1+n^2)^{s'/2}\hat{\phi}_1*\hat{\phi}_2*\hat{\phi}_3$ is in $\ell^2$
provided $s'r'<-1$ where $\frac{1}{r'}+\frac{1}{r}= \frac{1}{2}$.
This is so if $s'<0$ ($r'$ is positive) and $s'<-\frac{1}{r'}=
\frac{1}{r}-\frac{1}{2}= \frac{3}{p'}-\frac{5}{2}<
3(s+\frac{1}{2})-\frac{5}{2}= 3s-1$. The special case with $s'=0,\,\,
s>1/3$ is an immediate consequence of the first inequality
(\ref{Sobolev1}).  \hfill \qed

\section{Estimates for the Non-linear Equations}

\subsection{Local Existence}

The Duhamel integral representation of the system equations for
$\uvec$, 
Eq.(\ref{DIFFEQ}), can be written
\begin{equation}\label{Duhamel}
      \uvec (t) = \int_0^t e^{(t-t')L_{o}}\left(\begin{array}{c}
 \frac{\mu}{B}\phi^3(t')dt'\\ \sqrt{2T}\,d\omega(t')
   \end{array}\right) +
   e^{tL_{o}}\uvec(0).
\end{equation}
Fix $s$ with $\frac{1}{6}<s<1$, and for $R>1$ let ${\cal
D}_{R}(\beta,t)$ be the set of functions defined
\begin{eqnarray}\label{DRdef}
{\cal D}_{R}(\beta,t) &\equiv& \left\{\uvec(\cdot)\in {\cal C}([0,t],
H^s)\,|\, \|\uvec(0)\|_{H^s}\leq
\beta\phantom{\sup_0Y}\right.\nonumber\\&&
\phantom{XXXXX}\left.{\rm and}\,\,\sup_{t'\leq t}\|\uvec(t')\|_{H^s}\leq
  R\beta \right\},                            
\end{eqnarray}
and let ${\cal F}_R(\beta,t)$ be the (probabilistic) event that the
Duhamel integral equation Eq.(\ref{Duhamel}) has a unique {\em strong}
solution in ${\cal D}_R(\beta,t)$.  We have the following local
existence result.
\begin{prop}\label{localprop}
  Assume $\frac{1}{6}<s<1$.  There exist constants $c_1$, $c_2$, $c_3$
  and $C$ such that if $\uvec(0)$ satisfies
  $\|\uvec(0)\|_{H^s}\leq\beta$, $R>3c_3$ and $t\leq
  c_1/(R^2\beta^2)$, then
\begin{equation}
P\{{\cal F}_R(\beta,t)\}\geq 1- C\exp{\left(-\frac{c_2R^2\beta^2}{t(1+t)^2}\right)}.
\end{equation}
  \end{prop}

Clearly, the sets ${\cal F}_R(\beta,t)$ are nested, ${\cal
F}_R(\beta,t_2)\subset {\cal F}_R(\beta,t_1)$ if $t_1\leq t_2$. The
event ${\cal F}_R(\beta)\equiv \cup_n {\cal F}_R(\beta,t/n)$ is the
event that $\uvec(\cdot)$ exists for some positive time, and, in this
time, is no bigger than $R\beta$. An immediate corollary of the
above proposition is that ${\cal F}_R(\beta)$ occurs with probability
one.
\begin{corollary}  For $s>\frac{1}{6}$, local existence of the solution 
$\uvec(\cdot)$ in $H^s$ holds almost surely, 
\begin{equation}
P\{{\cal F}_R (\beta)\}= 1.
\end{equation}
\end{corollary}

\noindent {\em Proof of Proposition \ref{localprop}}:  We have 
\begin{eqnarray}\label{poil}
\lefteqn{\left\|\int_0^te^{(t-t')L_{o}}\left(\begin{array}{c}
 \frac{\mu}{B}\phi^3(t')dt'\\ 0 \end{array}\right)\right\|_{H^s}\leq
 c_3t\mu\sup_{t'\leq t}\|\phi^3(t')\|_{H^{s-1}}}\nonumber\\ &\leq& c
 \mu t\sup_{t'\leq t}\|\uvec(t')\|_{H^s}^3 \leq c\mu t (R\beta)^3 < \frac{1}{3}R\beta,
\end{eqnarray}
for a suitable constant $c$. Here, we have used Lemma \ref{lemmaBS2} and the 
Sobolev inequality Ineq. (\ref{Sobolev3}) of Lemma \ref{Sob} to
estimate $\|\phi^3\|_{H^{s-1}}$, assuming that $s>1/6$ and using $s-1<
\min(0,3s-1)$. Also, we have chosen $t < ( 3 c \mu R^2 \beta^2)^{-1}$.
Furthermore 
\begin{equation}\label{inject}
\left\|\int_0^te^{(t-t')L_{o}}\left(\begin{array}{c}
 0\\ \sqrt{2T}\,d\omega(t') \end{array}\right)\right\|_{H^s}< \frac{1}{3}R\beta\,,
\end{equation}
using that the left side Ineq.(\ref{inject}) is bounded by $c(1+t)
\sup_{t'\leq t}|\omega(t')|$ for a suitable constant $c$ (cf. the proof of
Lemma \ref{lemmasigma0}); this condition holds with probability exceeding
\begin{eqnarray}\label{probest6}
P\left\{\sup_{t'\leq t}|\omega(t')|\leq \frac{R\beta}{3c(1+t)}\right\}
&\geq & 1- C\exp{\left(-\frac{c_2R^2\beta^2}{t(1+t)^2}\right)} 
\end{eqnarray}
for suitable constants $c_2,\,\,C$.  
Finally, since $R>3c_3$  we have $\| e^{tL_{o}}\uvec(0)\|_{H^s}\leq c_3\beta\leq
\frac{1}{3}R\beta$. Together with Ineqs. (\ref{poil},\ref{inject}) this implies that the right side of the 
Duhamel integral equation Eq.(\ref{Duhamel}) is
a map of ${\cal D}_{R}(\beta,t)$ into itself. 

It remains to check that the right side of the Duhamel equation
Eq.(\ref{Duhamel}) is contractive for  small $t$.  But
clearly for two functions $\uvec_1,\,\, \uvec_2\in {\cal
D}_{R}(\beta,t)$, with real field parts $\phi_1$ and $\phi_2$
respectively, and for $t_1\leq t$,
\begin{eqnarray}
    \lefteqn{\left\|\int_0^{t_1}e^{(t_1-t')L_{o}}\left(\begin{array}{c}
 \frac{\mu}{B}(\phi_1^3-\phi_2^3)(t')\\ 0
 \end{array}\right)dt'\right\|_{H^s}}\nonumber\\ 
 &\leq&
 c_3\mu t\sup_{t'\leq
 t}\|(\phi_1^2+\phi_1\phi_2+\phi_2^2)(\phi_1-\phi_2)(t')\|_{H^{s-1}}
 \nonumber\\ &\leq& 3 c \mu t(R\beta)^2\sup_{t'\leq
 t}\|(\uvec_1-\uvec_2)(t')\|_{H^s} 
 \end{eqnarray}
by Ineq.(\ref{Sobolev3}) of Lemma \ref{Sob}, for a suitable
constant $c$.  Thus the Duhamel integral is a contraction for
$t<  1/(3c \mu R^2\beta^2)$. 

In summary, if $t< c_1/\mu R^2\beta^2)$, for a suitably small constant $c_1$, and if 
the stochastic integral estimate Ineq.(\ref{inject}) holds, the right side of the Duhamel
expression maps ${\cal D}_{R}(\beta,t)$ into itself  and 
it is a contraction, so that by the contraction mapping theorem, 
Eq.(\ref{DIFFEQ}) has a unique strong solution in ${\cal D}_{R}(\beta,t)$.  
Ineq.(\ref{inject}) holds with probability at least that given in Ineq.(\ref{probest6}).  \hfill \qed

\subsection{Global Existence}

Finally, in this section we provide a {\em global estimate} for the
non-linear stochastic evolution $\uvec(t)$.  Following Bourgain's methods
 for the non-linear wave equation \cite{Bo1}, we set
\begin{equation}
\tilde{\uvec}_{N} (t) = (\tilde{u}_N(t), \tilde{r}(t))=(( u(t)- P_{> N}u_o(t)), (r-r_{o}) (t)),
\end{equation}
with the positive integer $N$ to be chosen later, $P_{> N}$ 
projection onto the Fourier modes $\{k: |k|> N \}$.  Here,
$\uvec(t)$ and $\uvec_{o}(t)$ (the linear evolution) are assumed to
begin with the same initial data, $\uvec(0)= \uvec_o(0)$, and are
driven by the same stochastic driving terms, so that they are {\em not}
independent: they are {\em coupled}.  The quantity $I_{N}(t)$ is
defined as it would be in the pure deterministic case,
\begin{equation}
      I_{N} (t)= {\cal E} (\tilde{\uvec}_{N}(t)),
      \end{equation}
with ${\cal E}$ the energy defined by equation (\ref{energydef}).
Set 
\begin{equation}
\theta_{*}\equiv \min\{\frac{1}{3}(4s-1), \frac{1}{3} (1-s),\gamma\}.
\end{equation}
We will assume below that $\alpha\in H^{\gamma}$ with $\gamma>0$, and that
  $1/3\leq s< 1$.

  Our main result is the following:
\begin{prop}\label{Prop3.2}
Let $\uvec(0)=\uvec_o(0)$, with
$\beta=\|\uvec(0)\|_{H^s}=\|\uvec_{o}(0)\|_{H^s}$.  Fix $R>1$, $\theta>0$
and $\delta>0$ so that $\theta+\delta< \theta_{*}$.  There exist
constants $c$, $C$ and an $N_{o}= N_{o} (R,\beta)$ and
$\tau=\tau(R,\beta)$, such that if $N\geq N_{o}$ and $t\leq \tau
N^{\delta}$, then
\begin{equation}
 P\left\{\sup_{t'\leq t} I_{N} (t')>R\beta^2N^{2 (1-s)} \right\}\leq C \exp{\left(-\frac{cN^{2\theta}}{t(1+t)^2}\right)}.
  \end{equation}
  \end{prop}

\noindent {\em Proof:} The stochastic
differential of $ I_{N}(t)$ is given by 
\begin{eqnarray}\label{eIeqn}
d I_{N}(t)&=&  \left(\Im \langle
B\tilde{u}_{N}, \mu(\Re u)^3-\mu(\Re \tilde{u}_{N})^3+r_{o} P_{\leq
N}\alpha\rangle - \tilde{r}^{2}\right.\nonumber\\
&&\left.\phantom{xxxxxxxx(\Re u_{o})^{3}}+\tilde{r}\Im\langle Bu_{o}, P_{\leq N}\alpha\rangle \right)dt.
\end{eqnarray}
In particular, there are no $d\omega$ or $d\omega^2=dt$ terms, hence
the differential is the same as if we were just considering a wave
equation with dissipation.  We proceed to estimate the terms on the
right side.

 We have that
\begin{eqnarray}
\|(\Re u)^3 (t) -(\Re\tilde{u}_{N})^3 (t)\|_2\!\!\!&\leq&\!\!\!\! c''\left(\|P_{> N}u_o (t)\|_2
\|\tilde{u}_N\|_{\infty}^{2}+ \|P_{> N}u_o (t)\|_6^3\right)\nonumber\\
 &\leq &\!\!\!\! c'\left(\|P_{> N}u_o (t)\|_2
\|\tilde{u} \|_{4}^{4/3}\|\tilde{u} \|_{H^1}^{2/3}+ \|P_{> N}u_o (t)\|_6^3\right)\nonumber\\
 &\leq & \!\!\!\!c\left(N^{-s}\|u_o (t)\|_{H^s}I_{N}(t)^{2/3}+\|u_o (t)\|_{H^s}^{3}\right)
 \end{eqnarray}
for suitable constants $c'', c', c$. Here we have used Sobolev
inequalities, first Ineq.(\ref{Sobolev2}), then Ineq.(\ref{Sobolev1}) of
Lemma \ref{Sob}.  The other factors in Eq.(\ref{eIeqn}) are 
readily estimated, and we get
\begin{eqnarray}
   dI_{N}(t)&\leq& c\left(N^{-s}\|u_o(t)\|_{H^s}I_{N}(t)^{7/6} \right. \nonumber\\
&&\!\!\!\!\left.+(|r_{o}(t)|+N^{1-\gamma-s}\|u_o(t)\|_{H^{s}}+\|u_{o}(t)\|_{H^s}^{3})I_N(t)^{1/2}\right)dt.
\end{eqnarray}

Now assuming that $I_{N} (t)\leq R\beta^2N^{2(1-s)}$ and $\|\uvec_o(t)\|_{H^s}\leq N^{\theta}$, with $0<\theta<\theta_*\equiv \min\{\frac{1}{3}(4s-1), \frac{1}{3} (1-s),\gamma\}$, we obtain  
  \begin{eqnarray}
 d I_{N}(t)   &\leq& N^{2(1-s)}\left(N^{\theta}{\cal O}(N^{\frac{1}{3}(1-4s)})+N^{3\theta} {\cal O}(N^{-(1-s)})+N^{\theta}{\cal O}(N^{-\gamma}) \right)dt\nonumber\\
&\leq& N^{2(1-s)} {\cal O}(N^{\theta-\theta_*})dt.
\end{eqnarray}
It follows that $I_{N} (t)< R\beta^2N^{2(1-s)}$ for $t\leq T$ with $T=
{\cal O} (N^{\delta})$, $\delta< \theta_*-\theta$, provided that in
this time interval, $\|\uvec_{o}(t)\|_{H^{s}}< N^{\theta}$. Said more precisely, given 
$\beta,\, R$, there exist an $N_{o} (\beta, R)$ and a $\tau(\delta, \theta, \beta,R)$ such that
for $N\geq N_{o}$ $I_{N} (t)$ remains less than $R\beta^2N^{2(1-s)}$ for a time
$t$, $0\leq t\leq \tau N^{\delta}$, {\em provided} that $\|\uvec_o(t)\|_{H^s}$ remains less than $N^{\theta}$ in this same time interval.

Thus we have that
\begin{eqnarray}
     \lefteqn{P\left\{\sup_{ t'\leq t} I_{N} (t')>R\beta^2N^{2 (1-s)} \right\}}\nonumber\\
&=&P\left\{\sup_{ t'\leq t} I_{N} (t')>R\beta^2N^{2 (1-s)}\,{\rm and}\, \sup _{ t'\leq t}\|\uvec_o(t')\|_{H^s} \geq N^{\theta}\right\}\nonumber \\
&\leq& P\left\{\sup_{ t'\leq t}\|\uvec_o(t')\|_{H^s} \geq N^{\theta}\right\}\nonumber \\
&\leq& C\exp{\left(-\frac{(N^{\theta}-c\beta)^2}{c^2t(1+t)^2}\right)}
\leq  C_{1}\exp{\left(-\frac{c_{1}N^{2\theta}}{t(1+t)^2}\right)}
\end{eqnarray}
for $ t\leq \tau N^{\delta}$, by Ineq.(\ref{sigma0bound}) of 
lemma \ref{lemmasigma0} and appropriate new constants $C_{1}$ and
$c_{1}$.  After renaming of constants and taking $N_o$ still larger so that 
$N_o^{\theta}> 2c\beta$, the proof of the proposition is complete.\hfill \qed 

The proposition \ref{Prop3.2} and the lemma \ref{lemmasigma0} give us a global bound:
\begin{corollary}\label{corollary}
Let  $\beta$, $\theta< \theta_*$, $R\geq 2$ be fixed, as in the above Proposition.   There exist constants, $c,\,\,C$ and $N_{1}= N_{1} (\beta, R,t)$, such that for any time $t$, and $N\geq N_1$,  
\begin{eqnarray}
P\left\{\sup_{t'\leq t }\|\uvec(t')\|_{H^s}>R\beta N^{1-s}\right\}
\,\leq\, C\exp{\left(-\frac{cN^{2\theta}}{t(1+t)^2}\right)}.
\end{eqnarray}
\end{corollary}

\noindent {\it Remark}: The same ideas, in particular estimating
$I_N(t)$ for any $N$, can be used to prove global existence almost
surely in the energy norm $s=1$, but we do not write
out precise statements here.\\

\noindent  {\em Proof:} We have that
\begin{eqnarray}\label{corest}
P\left\{\sup_{t'\leq t }\|\uvec(t')\|_{H^s}>R\beta N^{1-s}\right\}  &\leq& 
P\left\{\sup_{t'\leq t }\|\tilde{\uvec}_{N}(t')\|_{H^1}>\frac{1}{2}R\beta N^{1-s}\right\}\nonumber\\ 
&&\!\!\!\!\!\!\!\!\!\!\!\!+ P\left\{\sup_{t'\leq t }\|\uvec_{o}(t')\|_{H^s}>\frac{1}{2} R\beta N^{1-s}\right\}.  
\end{eqnarray}
Now the first probability on the right side is bounded by 
\begin{eqnarray}
P\left\{\sup_{t'\leq t}I_{N} (t')>\frac{1}{2}R^{2}\beta^{2} N^{2 (1-s)}\right\}  \,\leq\, C\exp{\left(-\frac{cN^{2\theta}}{t(1+t)^2}\right)}
\end{eqnarray}
for $N>N_o$ by Proposition \ref{Prop3.2}. 
The second probability on the right side of Ineq.(\ref{corest}) is
bounded by the estimate given in Lemma \ref{lemmasigma0}, with
$\lambda= \frac{1}{2}R\beta N^{1-s}>>{\cal O} (N^{\theta})$.
Thus this probability is negligible compared to the first term on the
right side of Ineq.(\ref{corest}).   Enlarging $C$ completes the proof
of the corollary. \hfill \qed 

\section{Large $k$ Cut-off Systems and an Equilibrium Invariant Measure}

\subsection{Convergence of Finite Dimensional Cut-off Systems}

  We consider a cut-off version of our system Eqs.(\ref{theeq}) where
we retain Fourier modes $\{k\}$ with $|k|\leq M$, $M$ a positive
integer.  Let $\uvec_M(t)= (u_M,r_M)(t)= (\phi_M+\frac{1}{B}\pi_M,
r_M)(t)$ denote a solution to the finite dimensional system 
\begin{eqnarray}\label{DIFFEQM}
   i\partial_tu_M(t)&=& Bu_M(t) +\frac{1}{B}P_{\leq M}
 (\mu\phi_M^3(t)+r_M(t)  \alpha)\nonumber\\ dr_M (t)&=& - (r_M
 (t)-\langle P_{\leq M}\alpha_{i},\pi_M(t)\rangle)dt
 +\sqrt{2T}d\omega(t)
\end{eqnarray}
for initial data $u_M(0)\in P_{\leq M}L^2$.The solution $\uvec_M(t)$
remains in $P_{\leq M}L^2$ and is clearly in $H^s$ (for any $s$),
since all Fourier coefficients $\hat{u}_{M,k}=0$ for $|k|>M$.  We remark
that under the same assumptions on the coupling functions $\alpha$,
the conclusions of the previous section, Propositions
3.1, \ref{Prop3.2} and Corollary \ref{corollary}, hold for solutions
$\uvec_M$ {\em uniformly} in $M$ with respective initial data
$\uvec_M(0)= P_{\leq M}\uvec(0)$ for an initial $\uvec(0)\in H^s$,
$s>1/3$.  In particular the arguments used there are equally valid for
the cut-off systems.

Fix $s$ and let ${\cal D}_{R}(\beta,t)$ be the set of continuous functions
defined in Eq.(\ref{DRdef}), in particular functions $\{\uvec \}$ bounded in the $H^{s}$-norm by $R\beta$ with $\|\uvec(0)\|_{H^s}\leq\beta$, and let
${\cal G}_{R}(\beta,t)$ be the probabilistic event defined
\begin{equation}\label{Gdef}
   {\cal G}_{R}(\beta,t)\equiv \{\uvec(\cdot),\,\,\uvec_M(\cdot)\in
   {\cal D}_{R}(\beta,t)\,\,{\rm for \,\,each\,\,}M\} 
\end{equation}
with $\uvec(\cdot)$ the solution to Eq.(\ref{theeq}) and
$\uvec_M(\cdot)$ the solution to  Eq.(\ref{DIFFEQM}).

\begin{prop}\label{procsconver}
 Fix $s>1/3$, a time $t>0$, and $s_o>s$. Then
  $\{\uvec_M(\cdot)\}$ converges strongly to $\uvec(\cdot)$ in $H^s$
  uniformly on
 ${\cal G}_R(\beta,t)\cap \{\uvec\,\,|
  \,\,\|\uvec(0)\|_{H^{s_o}}\leq \beta\}$. 
\end{prop}

\noindent {\em Proof}: For notational convenience we will replace the
$t$ of the Proposition statement by $t_1$, and work on the time
interval $t\in [0,t_1]$. We will assume that in this time interval we
have the a priori bounds $\|\uvec(t)\|_{H^s}\leq R\beta$ and
$\|\uvec_M(t)\|_{H^s}\leq R\beta$, for all $M$ and $t$.  Given
these bounds, we will actually show the stronger result that $P_{\leq
  M}\uvec-\uvec_M\rightarrow 0$ strongly in $H^1$ and
$P_{>M}\uvec\rightarrow 0$ strongly in $H^s$, uniformly in $t$ and the
initial data with $\|\uvec(0)\|_{H^{s_o}}\leq \beta$. 

 The quantities $\uvec$ and $\uvec_M$ satisfy the
respective Duhamel relations 
\begin{equation}\label{Du}
\uvec(t)=\int_0^t e^{(t-t')B_o}\left(\begin{array}{c}
    \frac{-i}{B}(\mu\phi^3+r\alpha)dt'\\
       \langle \alpha,\pi\rangle dt'+\sqrt{2T}d\omega(t')
\end{array}\right) + e^{tB_o}\uvec(0)
\end{equation}
and, for $\uvec_M$ with initial data $P_{\leq M}\uvec(0)$,
\begin{equation}\label{Dum}
\uvec_M(t)=\int_0^t e^{(t-t')B_o}\left(\begin{array}{c}
    \frac{-i}{B}P_{\leq M}(\mu\phi_M^3+r_M  \alpha)dt'\\
       \langle \alpha,\pi_M\rangle dt'+\sqrt{2T}d\omega(t')
\end{array}\right) + e^{tB_o}P_{\leq M}\uvec(0).
\end{equation}
We proceed to estimate $(P_{\leq M}\uvec-\uvec_M)(t)$ in the
$H^1$-norm, for $0\leq t\leq t_1$. 

From the above integral formulae
Eqs.(\ref{Du},\ref{Dum}) one sees that there will be contributions to
$(P_{\leq M}\uvec-\uvec_M)(t)$ coming from integrals involving the
non-linearity $\phi^3-\phi_M^3$, $(r-r_M)\alpha$, and
$\langle\alpha,(\pi-\pi_M)\rangle$.  We first consider the integral of the
non-linearity and obtain a contribution to $(P_{\leq
  M}\uvec-\uvec_M)(t)$
\begin{eqnarray}\label{nl1}
\lefteqn{\left\| \int_0^t e^{(t-t')B_o}\left(\begin{array}{c}
    \frac{-i\mu}{B}P_{\leq M}(\phi^3-\phi_M^3)dt'\\
       0
\end{array}\right)\right\|_{H^1}}\\
&\leq&\!\!\!\! c
(R\beta)^2\!\!\int_0^t\!\!\|P_{>M}\uvec(t')\|_{H^s}\,dt'+c(R\beta)^2\!\!\int_0^t\!\!\|(P_{\leq M}\uvec-\uvec_M)(t')\|_{H^1}\,dt'\nonumber
\end{eqnarray}
for a suitable constant $c$ by the Sobolev inequality
Ineq.(\ref{Sobolev3}) with $s'=0,\,\, s>1/3$.  Next, the contribution to
$(P_{\leq M}\uvec-\uvec_M)(t)$ from the integral of $(r-r_M)\alpha$ is
bounded by
\begin{equation}\label{nl2}
    \left\| \int_0^t e^{(t-t')B_o}\left(\!\!\begin{array}{c}
    \frac{-i(r-r_M)}{B}P_{\leq M}\alpha \,dt'\\
       0
\end{array}\!\!\right)\right\|_{H^1}\!\!\!\!\leq c\|\alpha\|_{2}\int_0^t\|P_{\leq M}\uvec-\uvec_M)(t')\|_{H^1}\,dt'.
\end{equation}
Finally, the contribution to
$(P_{\leq M}\uvec-\uvec_M)(t)$ from the integral of $\langle
\alpha, (\pi-\pi_M)\rangle$ is bounded by
\begin{eqnarray}\label{messy}
&&\|P_{>M}\alpha\|_2 \int_0^t \|P_{>M}\tilde{\uvec}(t')\|_{H^1}\,dt'
+\left| \int_0^t
  e^{-(t-t')}\langle \alpha,B P_{>M}
  \uvec_o(t')\rangle\,dt'\right|\nonumber\\
&&\quad \quad \quad + \|\alpha\|_2\int_0^t\|(P_{\leq M}\uvec-\uvec_M)(t')\|_{H^1}\,dt', 
\end{eqnarray}
where $\uvec_o$ is the solution to the linear equations (with
dissipation and noise), and $\tilde{\uvec}= \uvec-\uvec_o$.  
The second of these integrals is bounded by 
\begin{eqnarray}
\|P_{>M}\alpha\|_2\left\|\frac{e^{tL_o}-e^{-t}}{L_o+1}\uvec(0)\right\|_{H^1}&\leq&
c \|P_{>M}\alpha\|_2\|\uvec(0)\|_2\nonumber\\
&\leq &  c \|P_{>M}\alpha\|_2\beta  
\end{eqnarray}
by Lemma \ref{BSlemma} (with $s= 0$).
Adding these contributions, Ineq.(\ref{nl1},\ref{nl2},\ref{messy}), and
using the last estimate, we arrive at 
\begin{eqnarray}\label{messy2}
&&\|(P_{\leq M}\uvec-\uvec_M)(t)\|_{H^1}\leq
c(\alpha,R\beta)\int_0^t \|(P_{\leq
  M}\uvec-\uvec_M)(t')\|_{H^1}\,dt'\nonumber\\
&&\quad\quad\quad+ c(R\beta)^2\int_0^t\|P_{>M}\uvec(t')\|_{H^s}\,dt' +
\|P_{>M}\alpha\|_2 \int_0^t
\|P_{>M}\tilde{\uvec}(t')\|_{H^1}\,dt'\nonumber\\
&&\quad\quad\quad\quad + cR\beta \|P_{>M}\alpha\|_2,  
\end{eqnarray}
where $c(\alpha, R\beta)$ is linear in $\|\alpha\|_2$ and
quadratic in $R\beta$.  

The inhomogeneous terms on the right side of this inequality
(\ref{messy2}), i.e., the second, third, and fourth terms, each go to
zero, $M\rightarrow\infty$ uniformly in $t$ and the data. {\em Second
term of (\ref{messy2})}: Consider the projection of the integral
formula of Eq.(\ref{Du}) above onto $P_{>M}H^s$. To control the non-linear
contribution  to $\|P_{>
  M}\uvec(t')\|_{H^s}$, we use 
\begin{eqnarray}
\|\frac{1}{B}P_{>M}\phi^3\|_{H^s}&=&\|P_{>M}\phi^3\|_{H^{s-1}}\nn\\
&\leq&
M^{s-s''}\|P_{>M}\phi^3\|_{H^{s''-1}}\nn\\
&\leq& M^{s-s''}\|\uvec\|_{H^{s}}^3\leq M^{s-s''}(R\beta)^3
\end{eqnarray}
by the Sobolev inequality (\ref{Sobolev3}), with $s''$ chosen, $ s<s''< 1$. 
The $\alpha$- contribution presents little difficulty and is ${\cal
  O}(M^{-\gamma}R\beta \|\alpha\|_{H^{\gamma}})$, while the
inhomogeneous term is estimated $\|P_{>M}e^{tB_o}\uvec(0)\|_{H^s}\leq
c_3 M^{s-s_o}\|P_{>M}\uvec(0)\|_{H^{s_o}}\leq c_3 M^{s-s_o}\beta$, by Lemma
 \ref{lemmaBS2}. {\em Third term of (\ref{messy2})}: We have that
\begin{eqnarray}
  \|P_{>M}\tilde{\uvec}(t)\|_{H^{1}}&\leq&\int_0^t
\left\| P_{>M}\left(\frac{\mu}{B}
\phi(t')^3+\frac{(r-r_{o})(t')}{B} \alpha\right) \right\|_{H^{1}}\,dt' 
\nonumber\\
&&\leq \int_0^t\|P_{>M}(\uvec(t')^{3})\|_2\,dt'
+ 2R\beta t\|P_{>M}\alpha\|_2.
\end{eqnarray}
Here, $P_{>M}(\uvec(t')^{3})= P_{>M}((P_{\leq [M/3]}\uvec(t')+P_{>
  [M/3]}\uvec(t'))^{3})$, where $[M/3]$ denotes the greatest integer
$\leq M/3$.  Expanding this out, one sees that terms containing a factor
$P_{> [M/3]}\uvec(t')$ go to zero uniformly for $M\rightarrow\infty$ as
  in our analysis of the second term of (\ref{messy2}), and
the term $P_{>M}(P_{\leq [M/3]}\uvec(t')^3)$ is identically zero.  The
$\alpha$-term is ${\cal O}(M^{-\gamma}R\beta\|\alpha\|_{H^{\gamma}})$. 
It follows that $\|P_{>M}\tilde{\uvec}(t)\|_{H^{1}}$ goes to $0$ uniformly.  
{\em Fourth
term of (\ref{messy2})}: This term, proportional to $\alpha$, is also
${\cal O}(M^{-\gamma}R\beta\|\alpha\|_{H^{\gamma}})$.

Thus Ineq.(\ref{messy2}), with each of its inhomogeneous terms going
to zero uniformly, $M\rightarrow\infty$, implies via Gronwall's
inequality that $(P_{\leq M}\uvec-\uvec_M)(t)$ goes to zero in $H^1$
uniformly in $t$ and the data.  Since $P_{>M}\uvec(t)$ goes to zero
uniformly in $H^s$ as we have seen above, we have that $\uvec_M(t)$
converges to $\uvec(t)$ uniformly in $H^s$, {\em provided} that
$\|\uvec(t)\|_{H^s}$ and $\|\uvec_M(t)\|_{H^s}$ stay less than
$R\beta$ for $t\leq t_1$, i.e., are in ${\cal G}_R(\beta,t_1)$, and the
data $\uvec(0)$ satisfies $\|\uvec(0)\|_{H^{s_o}}\leq \beta$. \hfill
\qed

\subsection{Equilibrium Invariant Measure}

We proceed now to show the existence of an invariant measure for
the complete system Eqs.(\ref{theeq}), but in
equilibrium where all temperatures are equal to a common $T$. Let
$\nu^0 $ be the Gaussian measure referred to in the introduction,
which at least formally is the invariant measure for the linear
problem with the $\int\phi^4$-term turned off.  With respect to
$\nu^0$, $\| u\|_{H^{s}}^2$ has finite expectation for $s<1/2$,
\begin{equation}\label{var}
  \int \| u \|_{H^{s}}^2 d\nu^0 = 2T\sum_k (1+k^2)^{s-1}<\infty,
\end{equation}
hence $\|\phi\|_{H^{s}}$ is finite, $\nu^0$-a.s. By  Sobolev
inequality (\ref{Sobolev1}) of Lemma \ref{Sob}, $\|\phi\|_4\leq c
\|\phi\|_{H^s}$ for $s>1/4$, so that as random variables
$\|\phi\|_4$ and $\|P_{\leq M}\phi\|_4$ are also finite $\nu^0$- a.s.
(Actually $\nu_0$ is supported on continuous functions, but we do not need this here.)
 Set
\begin{eqnarray}
  d\nu_M(\pi,\phi)&=&
 Z_M^{-1}\exp{\left(-\frac{\mu}{4T}\int\phi_M^4\,dx\right)}
d\nu^0(\pi,\phi)\nonumber\\
 d\nu(\pi,\phi)&=&
 Z^{-1}\exp{\left(-\frac{\mu}{4T}\int\phi^4\,dx\right)}d\nu^0(\pi,\phi)
\end{eqnarray}
with appropriate normalizations
\begin{equation}
  Z_M=
  \int\exp{\left(-\frac{\mu}{4T}\int\phi_M^4\,dx\right)}\,d\nu^0,\,\,\,Z=
  \exp{\left(-\frac{\mu}{4T}\int\phi^4\,dx\right)}\,d\nu^0
\end{equation}
and $\phi_M= P_{\leq
M}\phi$. Since $\|\phi\|_4$ and $\|P_{\leq M}\phi\|_4$ are finite
a.s., the Radon-Nikodym factors
$\exp{\left(-\frac{\mu}{4T}\int\phi_M^4\,dx\right)}$ and
$\exp{\left(-\frac{\mu}{4T}\int\phi^4\,dx\right)}$ are bounded and
positive a.s., and the normalizations $Z$ and $Z_M$ are positive.
The measures $\nu_M$ and $\nu$ are absolutely continuous with respect
to $\nu^0$.

The semigroup associated with the cut-off system Eq.(\ref{DIFFEQM})
acts invariantly on functions $\{ f(\uvec)\}$ of the form $f(\uvec)=
g(\hat{u}_{-M}, \dots,\hat{u}_{M},r)$, with $g$
integrable. The measure $\nu_M$ is an invariant measure for this
semigroup, as can be checked by computing the generator of the process
and showing that its adjoint annihilates $\nu_M$; we leave this
exercise to the reader.

We also have that $\lim_{M\rightarrow\infty}Z_M= Z$. This is the case
by the bounded convergence theorem: Clearly the exponentials are
bounded by one, and
\begin{equation}
\left|\,\|\phi\|_4-\|\phi_M\|_4 \right|\leq \|\phi-\phi_M\|_4\leq
c\|\phi-\phi_M\|_{H^s}\rightarrow 0,\,\, M\rightarrow\infty\,\, {\rm a.s.}
\end{equation}
 so that
\begin{equation}
\exp{\left(-\frac{\mu}{4T}\int\phi_M^4\,dx\right)}\rightarrow
\exp{\left(-\frac{\mu}{4T}\int\phi^4\,dx\right)},\,\, {\rm a.s}.
\end{equation}

Let $f$ be a function in the norm closure $\bar{\cal X}$ of functions
depending continuously on only a finite number
 of modes,
\begin{equation} {\cal X}\equiv \cup_M \{f\,\,|\,\, f=
g(\hat{u}_{-M}, \dots,\hat{u}_{M}, r), \,\,\,g\,\,{\rm bounded\,\,
continous}\}.
\end{equation}
Then again we have by bounded convergence
that

\begin{equation}
 \int f\exp{\left(-\frac{\mu}{4T}\int\phi_M^4\,dx\right)}d\nu^0\rightarrow
\int f \exp{\left(-\frac{\mu}{4T}\int\phi^4\,dx\right)}d\nu^0
\end{equation}
 and so
\begin{equation}
 Z_M^{-1}\int f\exp{\left(-\frac{\mu}{4T}\int\phi_M^4\,dx\right)}d\nu^0 
 \rightarrow  Z^{-1} \int f
\exp{\left(-\frac{\mu}{4T}\int\phi^4\,dx\right)}d\nu^0.
\end{equation}
Thus, $\nu_M$ converges to $\nu$ in a weak-$\star$ sense.

For later use, we also note a kind of tightness for the measures
$\{\nu_M \}$; for $s<1/2$,
\begin{eqnarray}\label{tightness}
   \int_{\{\|\uvec\|_{H^s}>\beta\}}\!\!\!\!\! d\nu_M&\leq& \frac{1}{Z_{M}}\int \frac{\|\uvec\|_{H^s}}{\beta}\exp{\left(-\frac{\mu}{4T}\int\phi_M^4\,dx\right)}d\nu^0\nn \\ 
&\leq& \frac{1}{\beta Z_{M}}\left(\int \|\uvec\|_{H^s}^{2}d\nu^0
\right)^{1/2}
\end{eqnarray}
which is arbitrarily small for $\beta$ large, uniformly in 
$M$, by Ineq.(\ref{var}).

Finally, we address the invariance of $\nu$.  Define the semigroups
\begin{equation}
  S^{t}f(\uvec )\equiv E_{\uvec}[f (\uvec (t))],\,\, S_{M}^{t}f(\uvec )\equiv
E_{\uvec}[f (\uvec_{M}(t))],
\end{equation}
where $f\in \bar{\cal X}$. (We will assume
here for definiteness that the Fourier modes $\hat{\uvec} _{M,k} (t)$
are simply constant in time for modes $|k|>M$.)

\begin{prop}\label{invariance} (Equilibrium case)
The measure $\nu$ is invariant with respect to the semigroup $S^{t}$
in the sense that for $f\in \bar{\cal X}$,
\begin{equation}
       \int S^{t}f\,d\nu= \int f\,d\nu.
\end{equation}
\end{prop}

\noindent {\em Proof}: Choose $s$, $1/3<s<1/2$, and let ${{\cal
    G}_R(\beta,t)}$ be the event defined in Eq.(\ref{Gdef}) (using the
$H^s$-norm).  By
Corollary \ref{corollary}, we have that for an initial $\uvec$ with
$\|\uvec\|_{H^s}\leq \beta$,
   \begin{equation}
     P_{\uvec}\{{{\cal
    G}_R^c(\beta,t)}\}\leq C\exp{\left(-\frac{c
    R^{2\theta/(1-s)}}{t(1+t)^2}\right)};
\end{equation}
we identify the $R$ here with $2N^{1-s}$ in the corollary statement (the
$R$ of the corollary being chosen equal to 2) and appropriately redefine the
constant $c$.  

Now let $s_o$ be chosen, with $s<s_o< 1/2$ and let $f(\uvec)$ be a
bounded function continuous in the $H^s$-norm of $\uvec$. (Such
functions are dense in $\bar{\cal X}$). For $\|\uvec\|_{H^{s_o}}\leq
\beta$ and any $\epsilon>0$,
\begin{eqnarray} 
\lefteqn{|S^{t}f (\uvec)-S^{t}_{M}f (\uvec)|}\nonumber\\&&\leq 
\left| E_{\uvec}[\chi_{{\cal G}_R(\beta,t)} (f(\uvec (t))-f (\uvec_M (t)
))]\right|+ 2\|f\|_{\infty}P_{\uvec}\{{\cal G}_{R}^{c} (\beta,t)
\}\nn\\ &&< \epsilon
\end{eqnarray} 
for $R$ and then M chosen sufficiently large, by the above probability
estimate, and by the uniform convergence of $\uvec_M$ to $\uvec$ on
${\cal G}_R(\beta,t)\cap \{\uvec\,|\,\|\uvec(0)\|_{H^{s_o}}\leq
\beta\}$, Proposition \ref{procsconver}.  Consequently, $S^{t}_{M}f
(\uvec)\rightarrow S^{t}f (\uvec)$, for $M\rightarrow\infty$ uniformly
in $\uvec$, $\|\uvec\|_{H^{s_o}}\leq\beta$.

Finally, 
\begin{eqnarray}\label{messy3}
\lefteqn{\left|\int S^{t}f d\nu-\int S_{M}^{t}f d\nu_M\right|}\nn\\
  &&\leq \left|\int S^{t}f (d\nu-d\nu_M)\right|+ \int \chi_{\{\|\uvec\|_{H^{s_o}}\leq \beta \}} \left|S^{t}f-S_{M}^{t}f\right|d\nu_M\nn \\
&& +2\|f\|_{\infty}\int _{\{\|\uvec\|_{H^{s_o}}> \beta \}}\!\!\!\!\!\! d\nu_M.
\end{eqnarray}
The first term on the right side goes to zero by weak-$\star$ convergence of
$\{\nu_M\}$ to $\nu$, the last term can be made arbitrarily small for
$\beta$ suitably large by tightness Ineq.(\ref{tightness}), and the
middle term then goes to zero by uniform convergence of $S^{t}_{M}f$
for $\|\uvec\|_{H^{s_{o}}}\leq \beta$.  Thus
\begin{equation}
   \int S^{t}fd\nu= \lim_{M\rightarrow\infty}\int S_{M}^{t}fd\nu_{M}  = 
   \lim_{M\rightarrow\infty}\int f d\nu_M = \int f d\nu
\end{equation}
by the above Ineq.(\ref{messy3}), by invariance of $\nu_M$ under
$(S^{t}_{M})^{*}$, and again by weak-$\star$ convergence of the
$\{\nu_M\}$.  This completes the proof of invariance of $\nu$ for
functions depending continuously on $\uvec $ with respect to the
$H^{s}$-norm and, by density, invariance for all $f\in \bar{\cal X}$. \hfill
\qed

\noindent {\em Remarks}: We emphasize that the question of ergodicity
for this equilibrium measure $\nu$ remains open, as does the existence
of non-equilibrium invariant measures for differing temperatures.\\

\noindent {\em Acknowledgments}: L.E.T. wishes to thank the Department
of Mathematics, U. Mass Amherst, for the hospitality extended to
him. L.R-B. was supported in part by NSF Grant DMS-0306540, L.E.T. by NSF Grant DMS-0245511.

\end{document}